\documentclass[twocolumn,showpacs,preprintnumbers,amsmath,amssymb]{revtex4}

\usepackage{epsfig} 
\usepackage{dcolumn}
\usepackage{bm}

\begin{document}

\preprint{APS/123-QED}

\title{Measurement of the radial matrix elements of the $6s \: ^2S_{1/2} \rightarrow 7p \: ^2P_{J} $ transitions in atomic cesium  }

\author{D. Antypas and D. S. Elliott}
\affiliation{%
  Department of Physics and School of Electrical and Computer Engineering \\ Purdue University, West Lafayette, IN  47907
}

\date{\today}

\begin{abstract}
We report measurements of the absorption strength of the cesium $6s \: ^2S_{1/2} \rightarrow 7p \: ^2P_{3/2} $ and the $6s \: ^2S_{1/2} \rightarrow 7p \: ^2P_{1/2} $ transitions at $\lambda = 456$ nm and 459 nm, respectively, in an atomic vapor cell.  By simultaneously measuring the absorption strength on the Cs D$_1$ line ($6s \: ^2S_{1/2} \rightarrow 6p \: ^2P_{1/2} $), for which the electric dipole transition moment is precisely known, we are able to determine the reduced dipole matrix elements for these two lines.  We report $\langle 7P_{3/2} || r || 6S_{1/2} \rangle  = 0.5780 \: (7) \: a_0$ and $\langle 7P_{1/2} || r || 6S_{1/2} \rangle = 0.2789 \: (16) \: a_0$, with fractional uncertainties of 0.12\% and 0.6\%, respectively.  Through these measurements, we can reduce the discrepancy between values of the tensor polarizability for the Cs $6s \: ^2S_{1/2} \rightarrow 7s \: ^2S_{1/2} $ transition that have been determined through two different methods.
\end{abstract}

\maketitle

\section{Introduction}
Precise measurements of the parity violating transition amplitude of the $6s \: ^2S_{1/2} \rightarrow 7s \: ^2S_{1/2} $ transition in atomic cesium~\cite{WoodBCMRTW97}, combined with accurate atomic structure calculations~\cite{BlundellJS90,BlundellSJ92,Derevianko00,DzubaF00,DzubaFG02,PorsevBD09,PorsevBD10,DzubaF12}, have provided a sensitive test of the Standard Model.  The laboratory determination of the weak-force-induced electric dipole transition moment $\mathcal{E}_{PNC}$ between the 6S and 7S states depends critically on accurate knowledge of the vector polarizability $\beta$ of the transition, since the laboratory measurements of Ref.~\cite{WoodBCMRTW97} yield the ratio $\mathcal{E}_{PNC}/\beta$.  There exist two independent determinations of $\beta$. The first, $\beta =26.957 \: (51) \: a_0^3$~\cite{DzubaF00}, is from a precision measurement of the ratio $M_1^{\rm hfs} /\beta$~\cite{BennettW99} ($M_1^{\rm hfs} $ is the off-diagonal contribution to the magnetic dipole transition moment), combined with an accurate calculation of $M_1^{\rm hfs} $~\cite{BouchiatG88,SavukovDBJ99,DzubaF00}. The second, $\beta=27.15 \: (11) \: a_0^3$~\cite{DzubaFG02} comes from an accurate measurement~\cite{ChoWBRW97} of the scalar to vector polarizability ratio $\alpha / \beta =  -9.905 \: (11) $, combined with a semi-empirical calculation of $\alpha$~\cite{DzubaFS97,VasilyevSSB02,DzubaFG02}.  $\alpha$, unlike $\beta$, can be calculated to high accuracy~\cite{DzubaFS97}. The two independent determinations of $\beta$ differ by $\sim 1.7 \sigma$, where $\sigma$ is the combined error in the difference. In this work, we reduce the discrepancy between these values of $\beta$ through a new laboratory measurement of the electric dipole transition matrix elements of the $6s \: ^2S_{1/2} \rightarrow 7p \: ^2P_{3/2} $ and the $6s \: ^2S_{1/2} \rightarrow 7p \: ^2P_{1/2} $ transitions.

The scalar polarizability $\alpha$ can be calculated as a sum over products of electric dipole matrix elements of the S states and intermediate P states~\cite{BouchiatB75}
\begin{eqnarray}\label{eq:alphaSumOverStates}
    \alpha & = & \frac{1}{6} \sum_n \left[ \rule{0mm}{5.5mm}  \langle 7S_{1/2} || r || nP_{1/2} \rangle  \langle nP_{1/2} || r || 6S_{1/2} \rangle  \right.     \nonumber \\
    & & \times \left( \frac{1}{E_{7S} - E_{nP_{1/2}}}  + \frac{1}{E_{6S} - E_{nP_{1/2}}} \right) \nonumber \\
    & & \rule{0mm}{5.5mm} - \langle 7S_{1/2} || r || nP_{3/2} \rangle  \langle nP_{3/2} || r || 6S_{1/2} \rangle \nonumber \\
    & & \left. \times \left( \frac{1}{E_{7S} - E_{nP_{3/2}}} +  \frac{1}{E_{6S} - E_{nP_{3/2}}} \right) \right].
\end{eqnarray}
$E_i$ represents the energy of state $|i \rangle$.  While the summation of Eq.~(\ref{eq:alphaSumOverStates}) extends over all $nP$ states, the major contributions come from the matrix elements involving the 6P and 7P states.  The primary contribution to the uncertainty in $\alpha$ is currently the $6S \rightarrow 7P_{3/2}$ matrix element, whose uncertainty is 0.8\%~\cite{VasilyevSSB02}.  One of the primary limitations on the precision of that measurement was the density of the Cs vapor in the cell.  In the measurements described in this work, we are able to reduce the uncertainty in this matrix element to 0.12\% by comparing the absorption strength of this transition to that of the D$_1$ line, whose transition moment is very well known~\cite{YoungHSPTWL94,RafacT98,RafacTLB99,AminiG03,DereviankoP02,BouloufaCD07}, and through this provide a more precise corrected value of the scalar polarizability $\alpha$.

The $6S \rightarrow nP_{1/2}$ matrix elements are also used in the theoretical determination of $\mathcal{E}_{PNC}$~\cite{PorsevBD09}, which is calculated as a summation over $nP_{1/2}$ states for all $n$, of the products of electric dipole matrix elements and weak-force induced matrix elements.  The 0.27\% uncertainty in $\mathcal{E}_{PNC}$ is dominated by the uncertainties of other matrix elements in this summation, however, and we find that our new value for the $6S \rightarrow 7P_{1/2}$ element has little direct impact on $\mathcal{E}_{PNC}$.

In the following section, we discuss our laboratory technique for measurements of the absorption strength of these two transitions.  In Section~\ref{sec:analysis}, we discuss our analysis of the absorption lineshapes which allow us to determine the reduced matrix elements.  Finally, we discuss a correction to the scalar polarizability $\alpha$, based on the new values of the reduced matrix elements, and from this a corrected value of the scalar polarizability $\beta$.  We conclude in Section~\ref{sec:conclusion}.

\section{Measurement}\label{sec:measurement}
\begin{figure}
  \includegraphics[width=8cm]{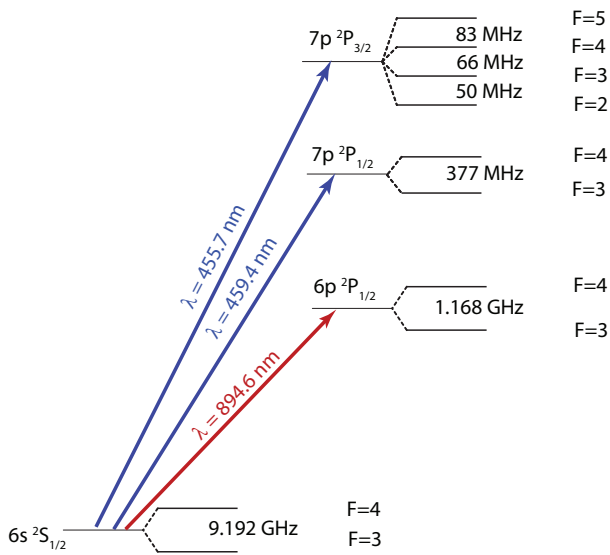}\\
  \caption{(Color on-line) Energy level diagram of atomic cesium, showing the levels relevant to these measurements.  Hyperfine splittings are taken from Ref.~\protect\cite{ArimondoIV77}.}
  \label{fig:EnergyLevels}
\end{figure}

We determine the radial matrix elements of the $6S \rightarrow 7P_{3/2}$ and $6S \rightarrow 7P_{1/2}$ transitions by measuring the absorption depth through a temperature-controlled atomic cesium vapor cell for a narrow band laser beam tuned to one of these transitions at 456 or 459 nm.  We compare the absorption by the cesium vapor at these wavelengths with the absorption of a 894 nm laser beam tuned to the $D_1$ resonance, employing laser beams carefully overlapped in the region of the atomic vapor, such that the beam path lengths are equal.  In this way, the need for precise knowledge of the Cs vapor density is avoided.  This method is similar to that used by Rafac and Tanner~\cite{RafacT98} in a measurement of the ratio of the $D_1$ and $D_2$ matrix elements.  Successive measurements of the Doppler-broadened absorption profile at the blue and IR wavelengths, repeated at different vapor densities, allow a determination of the $6s \: ^2S_{1/2}  \rightarrow 7p \: ^2P_{3/2} $ matrix element at $\lambda$ = 456 nm, $\langle 7P_{3/2} || r || 6S_{1/2} \rangle$, and the  $6s \: ^2S_{1/2}  \rightarrow 7p \: ^2P_{1/2} $ matrix element at $\lambda$ = 459 nm, $\langle 7P_{1/2} || r || 6S_{1/2} \rangle$, relative to the matrix element of the $6s \: ^2S_{1/2}  \rightarrow 7p \: ^2P_{1/2} $ transition at $\lambda$ = 894 nm.  The energy level diagram of Fig.~\ref{fig:EnergyLevels} shows details of the relevant states for these measurements.

\begin{figure}
  \includegraphics[width=8cm]{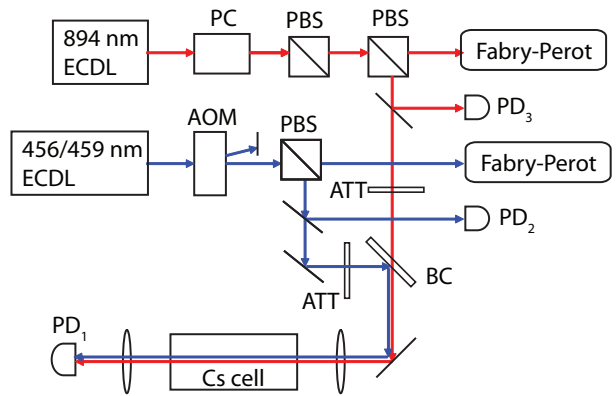}\\
  \caption{(Color on-line) Simplified schematic of the experimental apparatus. Abbreviations: ECDL, external cavity diode laser; PC, Pockels cell; PBS, polarizing beam splitter; PD, photodiode; AOM, acousto-optic modulator; ATT, attenuator; BC, beam combiner.  }
  \label{fig:setup}
\end{figure}

We show the experimental configuration in Fig.~\ref{fig:setup}.
The laser light required for these measurements is produced by two external cavity diode lasers (ECDL), each generating approximately 10 mW of optical power.  Through combined PZT and injection current scanning, these systems can be frequency scanned without mode-hopping through approximately 10 GHz, a range sufficient to obtain the Doppler-broadened spectra of each of the transitions.  During a frequency scan, the output power of each laser exhibits a $\sim$10\% variation.  We stabilize the optical power of the blue beam using photodiode PD$_2$ and an acousto-optic modulator, and the power of the IR laser beam using the PD$_3$ signal to control the voltage applied to a Pockels cell.  The magnitude of the residual power fluctuations after the power stabilizers is 0.1\% or less.  We measure a slight nonlinearity in the frequency scans for each ECDL by recording the transmission through a Fabry-Perot spectrum analyzer ($\sim$1.5 GHz FSR) and determining the position of the successive transmission peaks.  We apply a correction to the frequency scan of each of the recorded absorption spectra.  Although the lasers always operate single mode, a small amount of broadband power in the wings of the power spectrum is present in their outputs. This background is approximately 0.5\% of the power output for the 894 nm laser.  For the blue laser, this power in the wings is $\sim$1.5\% when operating at 456 nm, and $\sim4$\% at 459 nm.  We measure these levels by observing the transmission through a second cell, periodically inserted in the path of the two beams, which is heated to a high temperature, such that the narrow-band power in the lasing mode is completely absorbed in the vicinity of the absorption resonance.  We subtract this background from the recorded spectra.

The Cs vapor cell is 6 cm long and fitted with 0.5$^{\circ}$ wedged windows, in order to avoid etalon effects for  the transmitted beams.  To further minimize etalon effects, the two beams are slightly diverging through the cell.  The cell has a 2 cm cold finger, whose temperature is controlled independently from that of the cell body. The latter is heated with a heat rope wrapped around the cell.  Layers of aluminum tape around the rope, covered by insulating material aid in making the cell temperature uniform and stable.  The temperature is monitored with a thermocouple placed in contact with the cell body.  The cold finger is kept inside an aluminum block, whose temperature is controlled with a thermo-electric cooler (TEC).  The cold finger temperature is monitored with another thermocouple.  In order to prevent cesium condensation in the cell, the cold finger temperature is always lower than that of the cell by 7-15 $^{\circ}$C.  Magnetic fields in the cell, if strong enough, could affect our measurements, so we are careful to wrap this heat rope in opposite directions so as to minimize the impact of the heater current.  The Zeeman shift due to the earth's magnetic field is much less than the Doppler broadened linewidth of the transitions, so we do not attempt to compensate for this.

The power incident on the cell is $\sim$150 nW in the blue beam ($\sim$15 nW in the IR beam), with a beam diameter of $\sim$2 mm.  We measure the power transmitted through the cell with a photodiode (Hamamatsu S1336-8BK), labeled PD$_1$ in Fig.~\ref{fig:setup}, whose photocurrent is amplified in a low noise transimpedance amplifier (40 M$\Omega$ gain, 1.1 kHz BW).  Following a second amplification stage (gain of 6), the signal is fed to a Labview-controlled data acquisition system.

We carry out a set of absorption measurements as follows: Initially, with both laser beams blocked, we record the photo detector output for a 10 sec interval and measure the overall electronic offset in the signal. Then, we unblock the 894 nm beam, insert the high temperature cell in its path, and measure the broadband power level of the beam.  We then repeat this for the blue (456 or 459 nm) laser beam.  Afterwards, we remove the high-temperature cell, and with the blue beam blocked, record
three (two) spectra at 894 nm for calibration of the 456 nm (459 nm) absorption measurements, each of duration 3 sec.  We then repeat the same process for the 456 (3 scans at each temperature) or 459 nm (4 scans at each temperature) resonances.  In this quick succession from one wavelength to the other, there is negligible Cs density drift.  We then increase the cell cold finger temperature, allow time for the density to stabilize (typically $\sim$50 min.), and repeat the process described above.

There are only a few sources of systematic errors in recording the absorption profiles.  One type of error is any effect (other than absorption by the atomic vapor) that can affect the optical power for the beams reaching the photo detector.  Examples of such variations are imperfect power stabilization of the laser beams, etalon effects introduced along the path of the beams, and the slight beam motion as we scan the laser frequency (resulting from the rotation of the ECDL diffraction grating).  With the use of apertures to clean up the beam profiles, and slight beam focusing onto the photo detector area, such power variations are in all spectra less than 0.3\%, but typically less than 0.1\%.  Furthermore, in our analysis of each absorption spectrum, we employ a fitting procedure that includes the effect of a varying background level.  Another systematic error is introduced if the blue and IR beam path lengths through the Cs cell are not perfectly matched.  For our carefully overlapped beams, we estimate that the path lengths of the two beams through the cell are equal to within 0.05\%.  Saturation of the observed resonances could also create problems.  We avoid these by keeping the 894 nm and 456/459 nm beam intensities well below the saturation level ($\sim$10$^{-4}$ I$_{\rm sat}$).

\section{Analysis}\label{sec:analysis}
\begin{figure}
  \includegraphics[width=7cm]{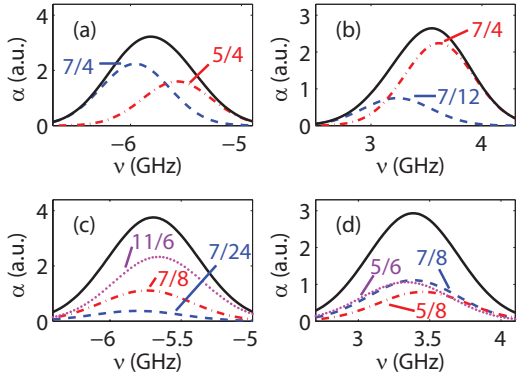}\\
  \caption{(Color on-line) Computed Doppler-broadened absorption lineshapes for the four different $6s \: ^2S_{1/2} F \rightarrow 7p \: ^2P_{J^{\prime}} $ absorption resonances.  These four lines represent transitions from (a) the F=4 level of the ground state to the $7p \: ^2P_{1/2} $ state at $\lambda = $ 459 nm, (b) the F=3 component of the ground state to the $7p \: ^2P_{1/2} $ state, (c) the F=4 component of the ground state to the $7p \: ^2P_{3/2} $ state at $\lambda = $ 456 nm, and (d) the F=3 component of the ground state to the $7p \: ^2P_{3/2} $ state.  The blue, dashed and red, dot-dashed lines in each represent transitions to the F$^{\prime}$ = 3 and F$^{\prime}$ = 4 levels, respectively.  The dotted magenta line is the $F=4 \rightarrow 5$ transition in (c), and the $F=3 \rightarrow 2$ transition in (d).  The solid black line in each shows the composite line shape, computed by summing the individual lines.  The frequency axis shows the frequency relative to the center of gravity of the transition.  The numerical factor displayed for each of the lineshapes is the angular momentum factor $q_{J,F \rightarrow J^{\prime},F^{\prime}}$ for that transition.}
  \label{fig:lineshapesfig}
\end{figure}

For each of the different absorption lines, we fit the transmission data to a lineshape of the form
\begin{equation}
T = T_{\rm max} (1-\exp\{-2 \alpha_{\rm abs} L\} ),
\end{equation}
where $T_{\rm max}$ is the power transmitted through the absorption cell when the laser is tuned far from resonance, $\alpha_{\rm abs}$ is the electric field absorption coefficient, and $L$ is the optical path length through the vapor cell.  The absorption coefficient $\alpha_{\rm abs}$ is of the form~\cite{RafacT98}
\begin{eqnarray}\label{eq:alpha}
\alpha_{\rm abs} &=& \frac{n \pi \alpha \omega }{\left( 2I+1\right) \left(2J+1\right) } \sqrt{\frac{M}{2 \pi k_B T}} \left| \langle J^{\prime} || \vec{r} || J \rangle \right|^2 \nonumber \\
  & & \times \sum_{F^{\prime}, m^{\prime}} \sum_{F, m} (-1)^{2(I+J)} \left( 2F^{\prime} + 1\right) \left( 2F + 1\right) \nonumber \\
  & & \times \left( \begin{array}{ccc} F^{\prime} & 1 & F \\ -m^{\prime} & 0 & m \end{array} \right)^2 \left\{ \begin{array}{ccc} J^{\prime} & F^{\prime} & I \\ F & J & 1 \end{array} \right\}^2 \\
  & & \times \int_{-\infty}^{\infty} \frac{\Gamma^{\prime} e^{-M v^2 / 2 k_B T} dv }{\left[\omega \left(1 - v/c \right) - \omega_{Fm \rightarrow F^{\prime}m^{\prime}} \right]^2 + \Gamma^{\prime 2}/4} \nonumber
\end{eqnarray}
for each transition from an initial state with hyperfine level of total angular momentum $F$ and projection $m$ to an excited state angular momentum $F^{\prime}$ and projection $m^{\prime}$.  In this expression, $n$ is the density of cesium atoms in the vapor cell, $\alpha$ is the fine structure constant, $\omega$ is the optical frequency of the laser field, and $\omega_{Fm \rightarrow F^{\prime}m^{\prime}}$ is the atomic transition frequency.  $M$, $k_B$, $T$ represent the atomic mass of the cesium atoms, the Boltzmann constant, and the cell temperature.
In the limiting case when the natural linewidth $\Gamma^{\prime}$ is much less than the Doppler-broadened linewidth $\Delta \omega_D = 4 \sqrt{\pi k_B T /M } / \lambda$, the integral in the final line of Eq.~(\ref{eq:alpha}), known as the Voight profile, becomes a Gaussian lineshape of width $\Delta \omega_D$ (full width at half maximum).  We show in Fig.~\ref{fig:lineshapesfig} computed absorption lineshapes for the four different $6s \: ^2S_{1/2} F \rightarrow 7p \: ^2P_{J^{\prime}} $ absorption resonances measured in this work.
We use the tables in Ref.~\cite{Steck} for the 6j and 3j symbols relevant for atomic cesium (with nuclear spin I=7/2), and display these angular momentum factors, which we call $q_{J,F \rightarrow J^{\prime},F^{\prime}}$, for each of the lineshapes in this figure.
Since the hyperfine splitting of the $7p$ states is smaller than the Doppler width, the individual peaks are not resolved.  The individual lines do contribute to the overall width of the composite lineshape, as well as a slight asymmetry in some cases (the F=3  $\rightarrow 7p \: ^2P_{1/2} $ line shown in Fig.~\ref{fig:lineshapesfig}(b), for example).

We determine the reduced matrix elements $\langle 7P_{J^{\prime}} || r || 6S_{1/2} \rangle  $ by measuring the relative strength of the absorption coefficient on two lines, $6s \: ^2S_{1/2} \rightarrow 7p \: ^2P_{J^{\prime}} $ at $\lambda$ = 456 nm ($J^{\prime}$ = 3/2) or $\lambda$ = 459 nm ($J^{\prime}$ = 1/2), and $6s \: ^2S_{1/2} \rightarrow 6p \: ^2P_{1/2} $ at $\lambda$ = 894 nm.  By comparing the ratio of the absorption coefficients at 894 nm and at 456 nm or 459 nm, we can eliminate the density $n$ of the Cs vapor from the analysis, and determine a relative measure of the transition moments.
Under these conditions, one can show that
\begin{equation}\label{eq:ratioalpha456}
\frac{\alpha_{F \rightarrow F^{\prime}}^{456} \: L}{\alpha_{3 \rightarrow 3}^{894} \: L} = \frac{ \langle 7P_{3/2} || r || 6S_{1/2} \rangle ^2 \: q_{1/2,F \rightarrow 3/2,F^{\prime}}}{\langle 6P_{1/2} || r || 6S_{1/2} \rangle ^2 \: \left(\frac{7}{12} \right)}
\end{equation}
and
\begin{equation}\label{eq:ratioalpha459}
\frac{\alpha_{F \rightarrow F^{\prime}}^{459} \: L}{\alpha_{3 \rightarrow 3}^{894} \: L} = \frac{ \langle 7P_{1/2} || r || 6S_{1/2} \rangle ^2 \: q_{1/2,F \rightarrow 1/2,F^{\prime}}}{\langle 6P_{1/2} || r || 6S_{1/2} \rangle ^2 \: \left(\frac{7}{12} \right).}
\end{equation}
The precision of the present measurements then depends upon the precision of the transition moment $\langle 6P_{1/2} || r || 6S_{1/2} \rangle $ for the D$_1$ transition.  For this we use the weighted average of the measurements of Refs.~\cite{YoungHSPTWL94,RafacTLB99,AminiG03,DereviankoP02,BouloufaCD07} to find $\langle 6P_{1/2} || r || 6S_{1/2} \rangle =  4.5062 \: (24) \: a_0$.  This value is in very good agreement with the {\it ab initio} results of $\langle 6P_{1/2} || r || 6S_{1/2} \rangle =  4.5064 \: (47) \: a_0$~\cite{DereviankoP02} and $  4.5093 \:  a_0$~\cite{PorsevBD10}.

\begin{figure}
  \includegraphics[width=7cm]{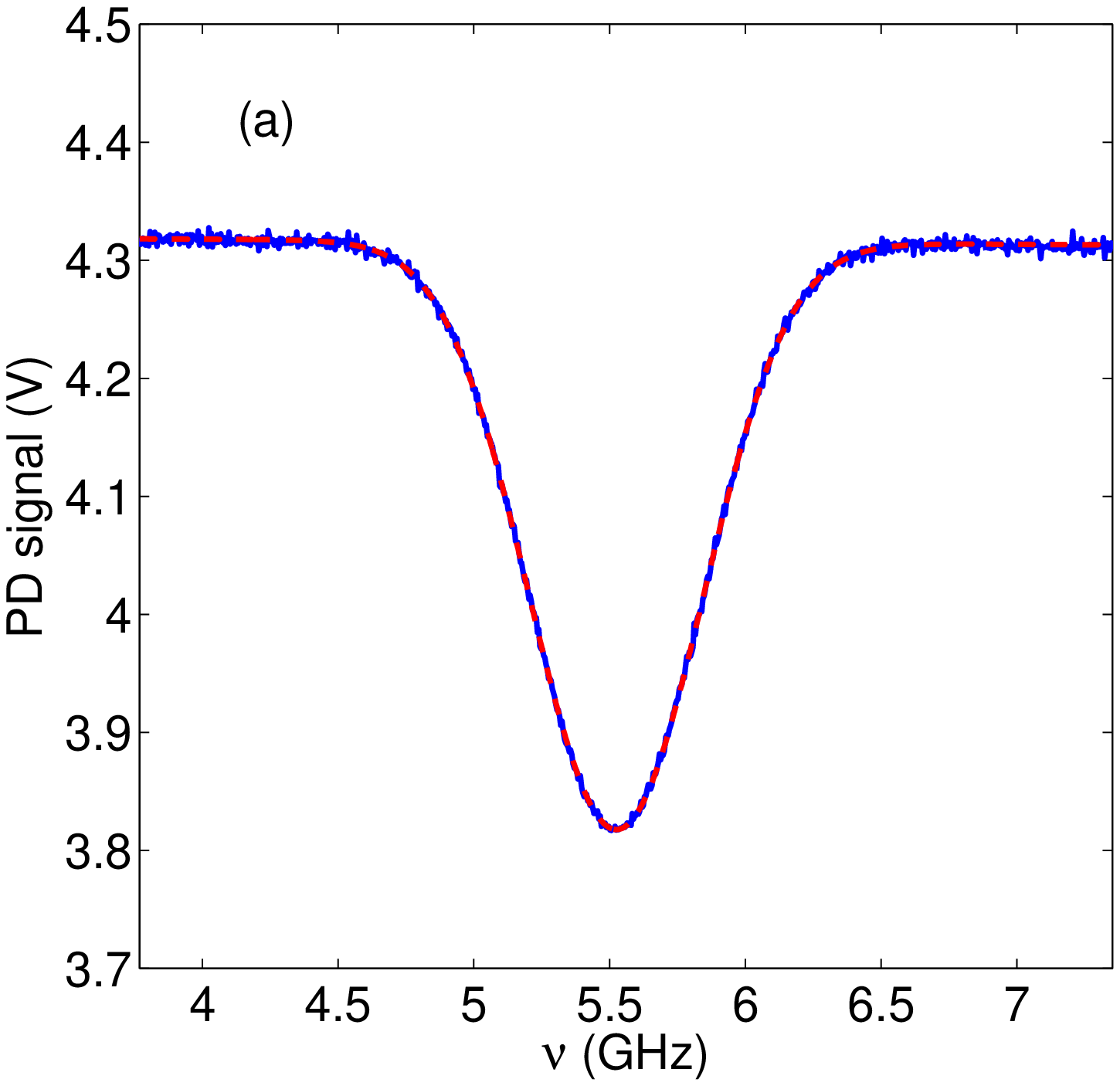}\\
  \includegraphics[width=7cm]{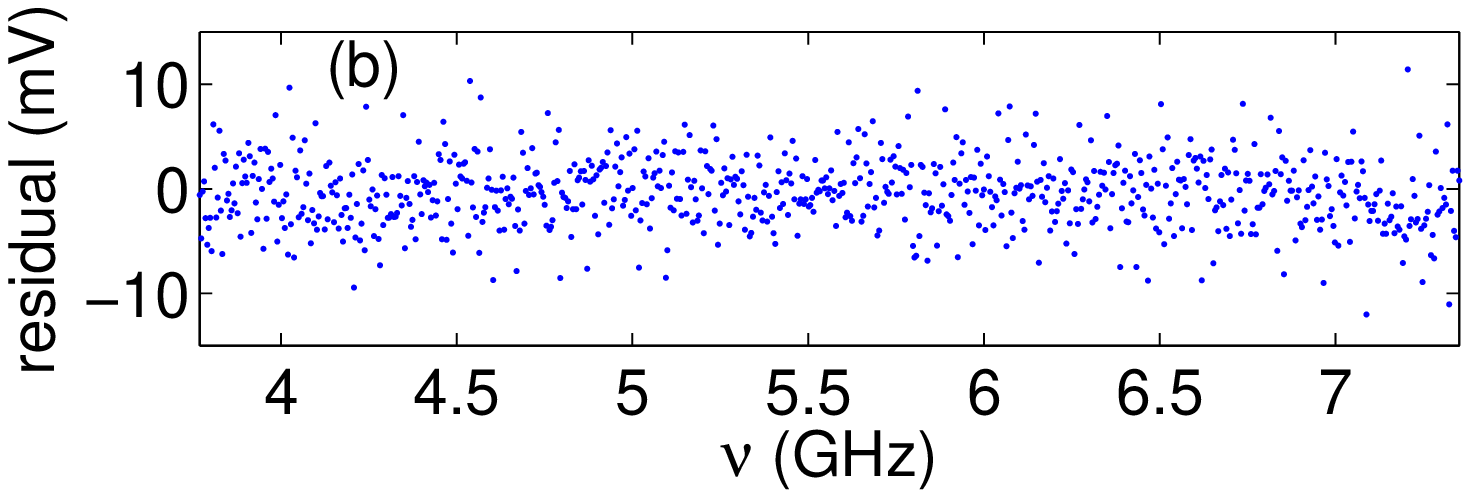}
  \caption{(Color on-line) (a) An example of an absorption spectrum for the transitions from the F=3 component of the ground state to the $7p \: ^2P_{3/2} $ state at $\lambda$ = 456 nm.  The blue data points represent the measured transmission of the laser beam through absorption cell, while the smooth red curve is the result of a least squares fit to the data.  (b) The residual between the transmitted power and the fitted curve.  }
  \label{fig:absdata456}
\end{figure}

We show an example of the absorption lineshape for the transitions from the F=3 component of the ground state to the $7p \: ^2P_{3/2} $ state at $\lambda$ = 456 nm in Fig.~\ref{fig:absdata456}(a).
The blue data points represent the measured transmission of the laser beam through the absorption cell, while the smooth red curve is the result of a least squares fit to the data.  In fitting the curve to the data, we adjust five parameters to minimize the root-mean-square deviation between the data and the fitted lineshape.  These parameters are: the frequency of the peak absorption, the minimum transmission (at line center), the level for full transmission (two terms, one to the left of line center, and a second to the right of line center, to account for slight variations in the transmitted power due to interference effects, etc.), and a scaling factor for the frequency scan.  The frequency difference between the hyperfine components, as well as their relative heights, however, are fixed.  We show the residual deviation between the measured transmission and the fitted curve in Fig.~\ref{fig:absdata456}(b).  This deviation is only a few millivolts in amplitude, relative to the measured signal of $\sim$4 V.  We can observe no systematic variation in this residual which might indicate a poor fit between the lineshape of Fig.~\ref{fig:lineshapesfig}(d) and the data.  The spectrum shown was taken at the highest cell temperature, corresponding to the highest cesium density.  We determine $\alpha_{3 \rightarrow 4}^{456} \: L$, the absorption coefficient on the $F=3 \rightarrow F^{\prime} = 4$ line of the $6s \: ^2S_{1/2} \rightarrow 7p \: ^2P_{3/2} $ transition, from the height of this absorption curve.

\begin{figure}
  \includegraphics[width=7cm]{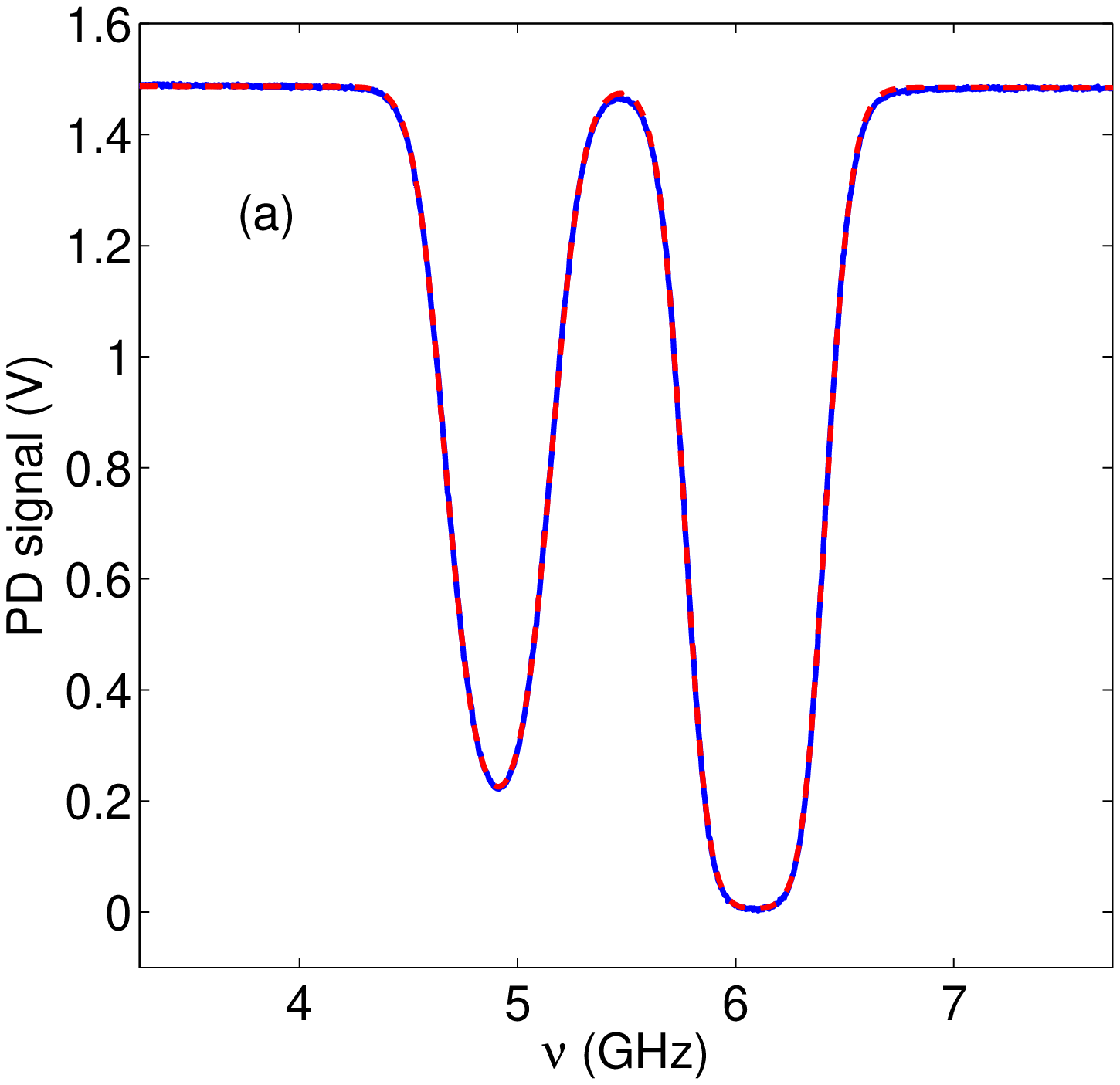}\\
  \includegraphics[width=7cm]{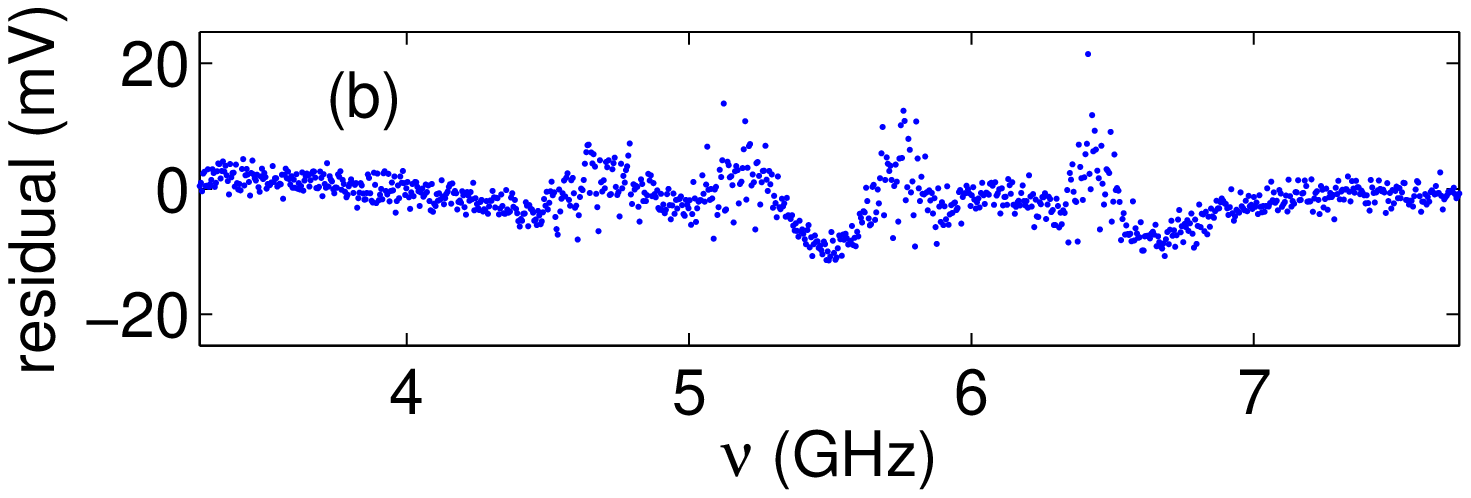}
  \caption{(Color on-line) (a) An example of an absorption spectrum for the transitions from the F=3 component of the ground state to the $6p \: ^2P_{1/2} $ line at $\lambda$ = 894 nm.  The blue data points represent the measured transmission of the laser beam through the absorption cell, while the smooth red curve is the result of a least squares fit to the data.  The weaker absorption peak on the left corresponds to the transition to the $F^{\prime}=3 $ level, while the stronger peak on the right is the transition to the $F^{\prime}=4 $ level. (b) The residual between the transmitted power and the fitted curve.     }
  \label{fig:absdata894}
\end{figure}

For each of the 456 nm transmission spectra, we also measure transmission spectra through the cell on the $D_1$ line, i.e. $6s \: ^2S_{1/2} \rightarrow 6p \: ^2P_{1/2} $ transition at 894 nm.  We show an example of this spectrum in Fig.~\ref{fig:absdata894}(a).
The blue data points and smooth red curve represent the measured and fitted transmission, respectively, as in the previous figure.  Here the two hyperfine lines, $F = 3 \rightarrow F^{\prime} = 3$ on the left, and $F = 3 \rightarrow F^{\prime} = 4$ on the right, are well resolved.  We adjust eight parameters in order to the determine the best fit to these data: the frequency at line center, the Doppler width, and the minimum transmission for each of the peaks, as well as the level of full transmission to either side of the resonances.  As a measure of the quality of the data, we monitor the fitted Doppler widths, the relative peak heights, and the relative peak areas.  The Doppler widths are consistent with the measured cell temperatures, while the ratios of the peak heights and areas on the two lines are consistent with 3, as expected on the basis of the angular momentum factors $q$ for these two lines.  We do observe some variation in these ratios with increasing cell density.  The ratio of peak heights is typically $\sim$2.98 or 2.99 at the lower densities, and decreases to 2.87-2.94 at the higher densities.  The ratio of peak areas shows a smaller variation with vapor density, in some cases increasing, in others decreasing.  Since the deviation of peak height ratio from 3 increases with increasing density, and since the absorption strength on the $F=3 \rightarrow F^{\prime}=3$ line is the weakest transition of the four D$_1$ lines [with an angular momentum factor $q$ of 7/12, relative to that of the other three lines of 5/4, 7/4 and 7/4, as shown in Fig.~\ref{fig:lineshapesfig}(a) and (b)], we use the absorption strength $\alpha_{3 \rightarrow 3}^{894} \: L$ of this particular line for calibration of the $6S \rightarrow 7P$ absorption measurements.

We show the residual between the data and the fitted transmission spectrum in Fig.~\ref{fig:absdata894}(b).  Here we can observe a small deviation between the transmitted power and the Doppler-broadened lineshape, with a systematic effect in the wings of each of the absorption line.  Still, the residual shown reaches a maximum value of only $\sim$10 mV, relative to 1.5 V signal at full transmission, and the fitted curves seem to capture the peak height of the absorption spectrum in each case.

\begin{figure}
  \includegraphics[width=7cm]{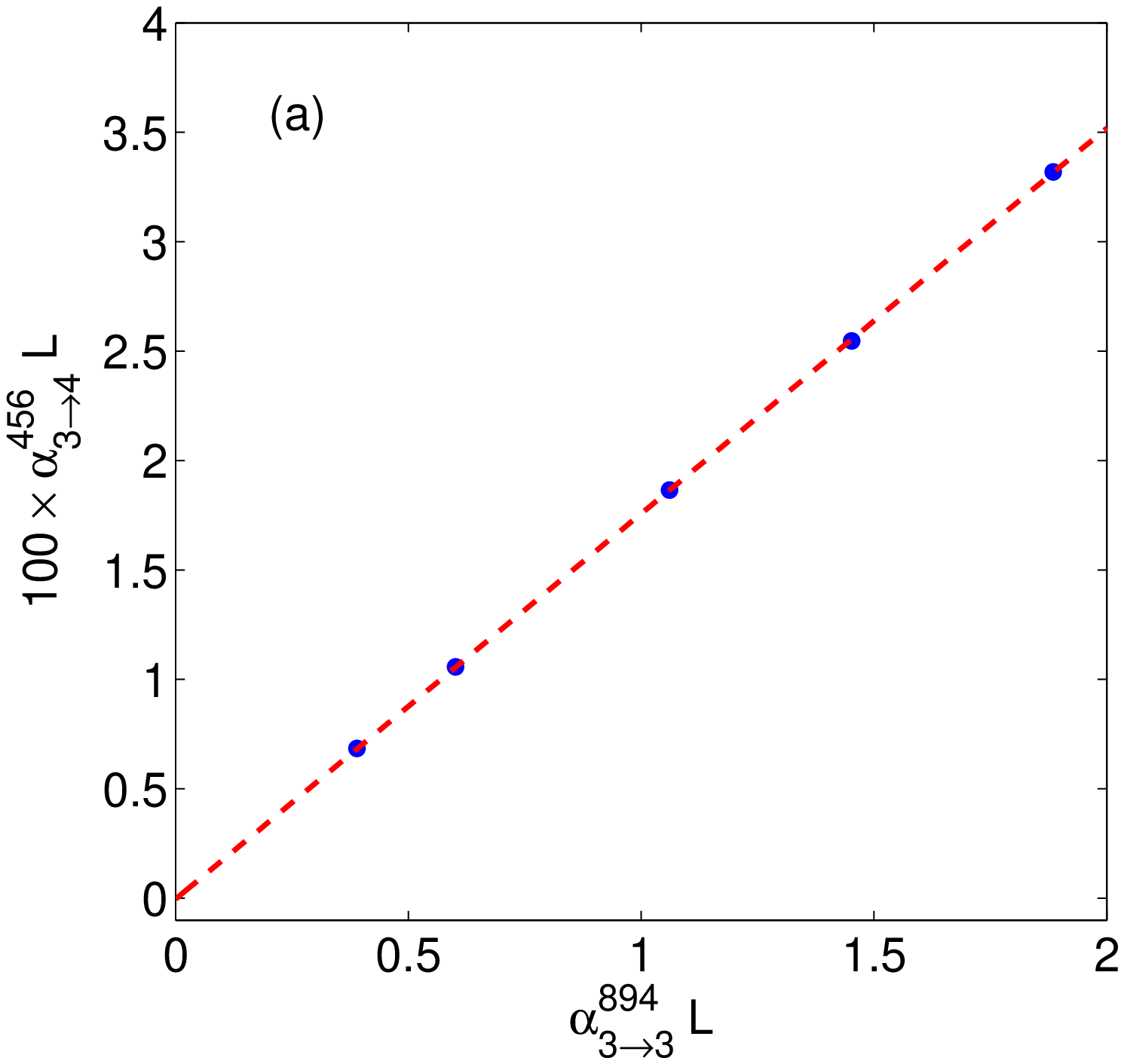}\\
  \includegraphics[width=7cm]{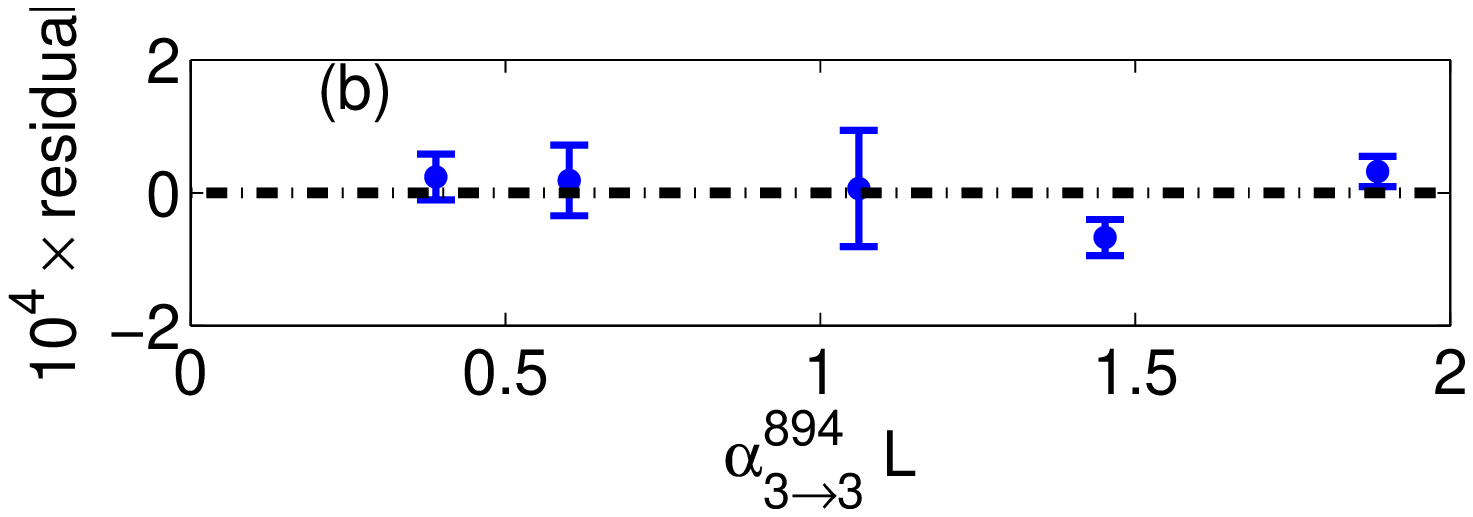}
  \caption{(Color on-line) (a) A plot the measured values of $\alpha_{3 \rightarrow 4}^{456} \: L$ at $\lambda$ = 456 nm versus $\alpha_{3 \rightarrow 3}^{894} \: L$ at $\lambda$ = 894 nm.  The red dashed line is the linear best fit to the data.  (b) The residual between the data points and the straight line fit.  The error bars on each residual indicate the standard deviation of the mean of the individual measurements.   }
  \label{fig:abs456vsabs894}
\end{figure}

We plot the measured values of $\alpha_{3 \rightarrow 4}^{456} \: L$ versus $\alpha_{3 \rightarrow 3}^{894} \: L$ in Fig.~\ref{fig:abs456vsabs894}(a).
In this plot each data point represents the average of three measurements of $\alpha_{3 \rightarrow 4}^{456} \: L$ at one of five different cell temperatures.  The red dashed line is the result of a linear least-squares fit to the data, with two adjustable parameters, the intercept $b$ and the slope $m$.  Ideally, one would expect the intercept to be zero, but several effects, such as an incomplete compensation for the optical power in the wings of the spectrum, a non-linear response of the photodiodes, or saturation of the transition, can result in a non-zero intercept.  The fitted intercept $b = -4 \: (6) \times 10^{-5}$ is consistent with 0, indicating that these effects are not significant in our measurements.  We show the residual of the data points from the best straight line fit in Fig.~\ref{fig:abs456vsabs894}(b).  The error bars on each residual indicate the standard deviation of the mean of the individual measurements.  The reduced $\chi^2$ for these data is 2.91.  As seen in these data, the residual error is smaller than the maximum absorption coefficient by $\sim$500, corresponding to only 0.1\% in the ratio of matrix elements.  We present the slope ($m$), intercept ($b$), and reduced $\chi^2$ of these data in Table~\ref{table:results456}.  The slope $m$ yields the reduced matrix element $\langle 7P_{3/2} || r || 6S_{1/2} \rangle = 0.5777 \: (7) \: a_0$ using Eq.~(\ref{eq:ratioalpha456}).  This uncertainty is derived from the statistical uncertainty of the fit of the slope of Fig.~\ref{fig:abs456vsabs894} only, and does not include the 0.05\% uncertainty in the $\langle 6P_{1/2} || r || 6S_{1/2} \rangle$ matrix element that we use for normalization.
\begin{table}
\begin{tabular}{|c|c|c|}
  \hline
      & $6s \: ^2S_{1/2} \: F=3  \rightarrow  $    & $6s \: ^2S_{1/2} \:  F=4  \rightarrow  $  \\
      &  \hspace{0.2in} $ 7p \: ^2P_{3/2} \: F=4$  &  \hspace{0.2in} $ 7p \: ^2P_{3/2} \: F=5$ \\ \hline \hline
  $m \times 10^{2}$ &  $1.761 \: (4)$   &  $5.187 \: (20)$  \\
  $b \times 10^{5}$   &  $-4 \: (6) $ &  $-13 \: (19) $  \\
  reduced $\chi^2$   & $2.91$   &  $1.87$  \\
  $q_{J,F \rightarrow J^{\prime},F^{\prime}}$  &  5/8  &  11/6  \\
  $\langle 7P_{3/2} || r || 6S_{1/2} \rangle $   & $0.5777 \: (7) $  &  $0.5789 \: (11) $  \\ \hline
\end{tabular}
\caption{Results of the least-squares fits to the data for the 456 nm absorption lines, including the slopes $m$, the intercepts $b$, the reduced $\chi^2$, and the reduced matrix element in atomic units $a_0$. }
\label{table:results456}
\end{table}

We also carry out similar measurements on the absorption line from the F = 4 component of the ground state to the $7p \: ^2P_{3/2} $ state.
We present the results of these measurements, using the absorption coefficient on the $F=4 \rightarrow F^{\prime} = 5$ line, $\alpha_{4 \rightarrow 5}^{894} \: L$, in the right column of Table~\ref{table:results456}.  The weighted average of these two values yields $\langle 7P_{3/2} || r || 6S_{1/2} \rangle  = 0.5780 \: (7) \: a_0$.  This fractional uncertainty of 0.12\% includes the 0.05\% uncertainty of the $\langle 6P_{1/2} || r || 6S_{1/2} \rangle$ matrix element.

\begin{figure}
  \includegraphics[width=7cm]{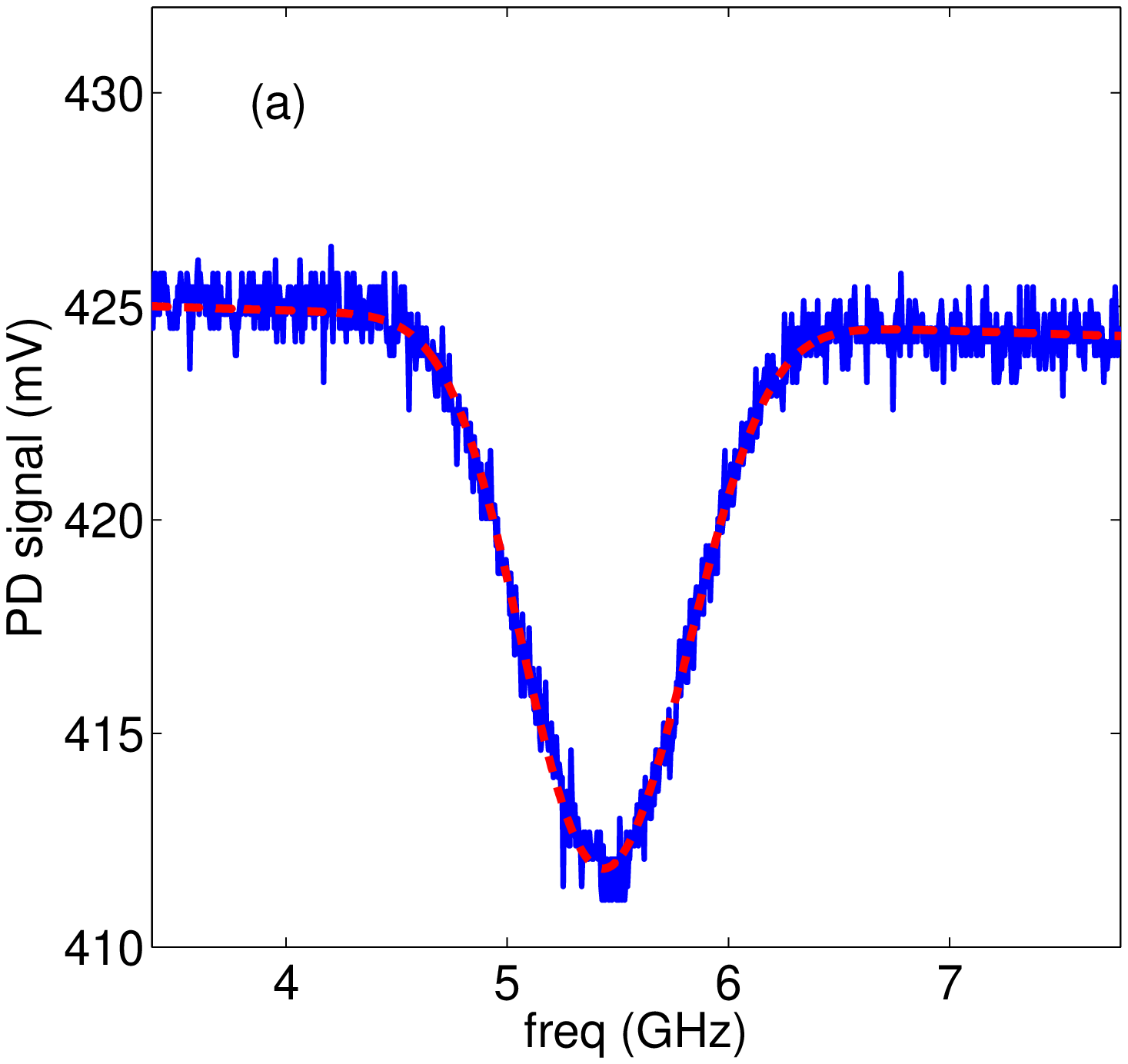}\\
  \includegraphics[width=7cm]{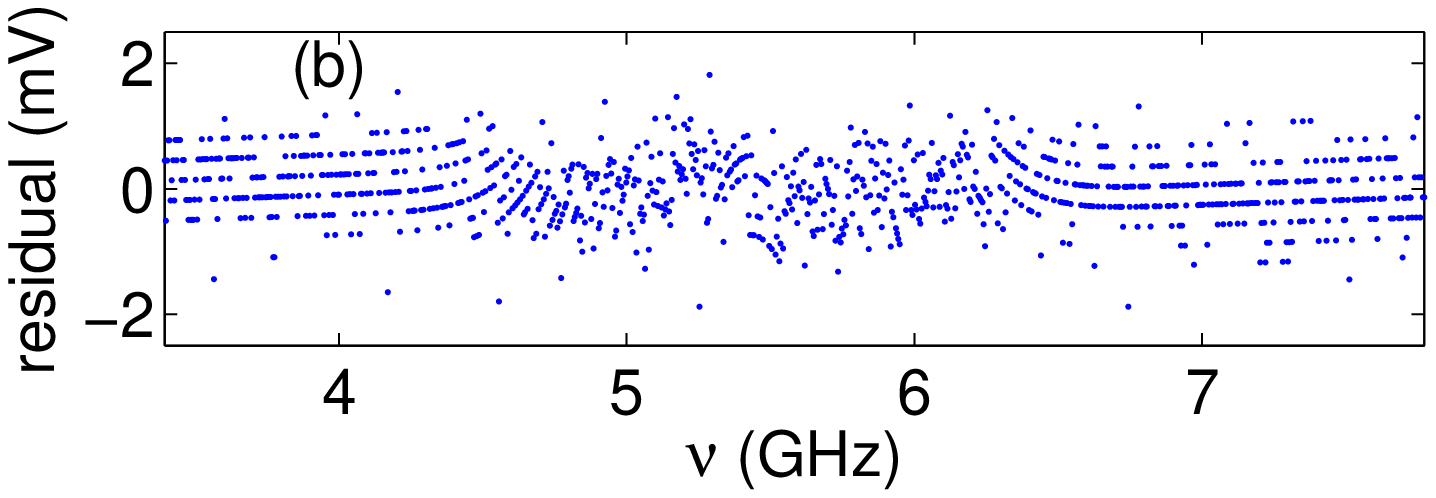}
  \caption{(Color on-line) Example of an absorption spectrum on the on the $6s \: ^2S_{1/2}  \rightarrow 7p \: ^2P_{1/2} $ line at 459 nm.    }
  \label{fig:absdata459}
\end{figure}

\begin{figure}
  \includegraphics[width=7cm]{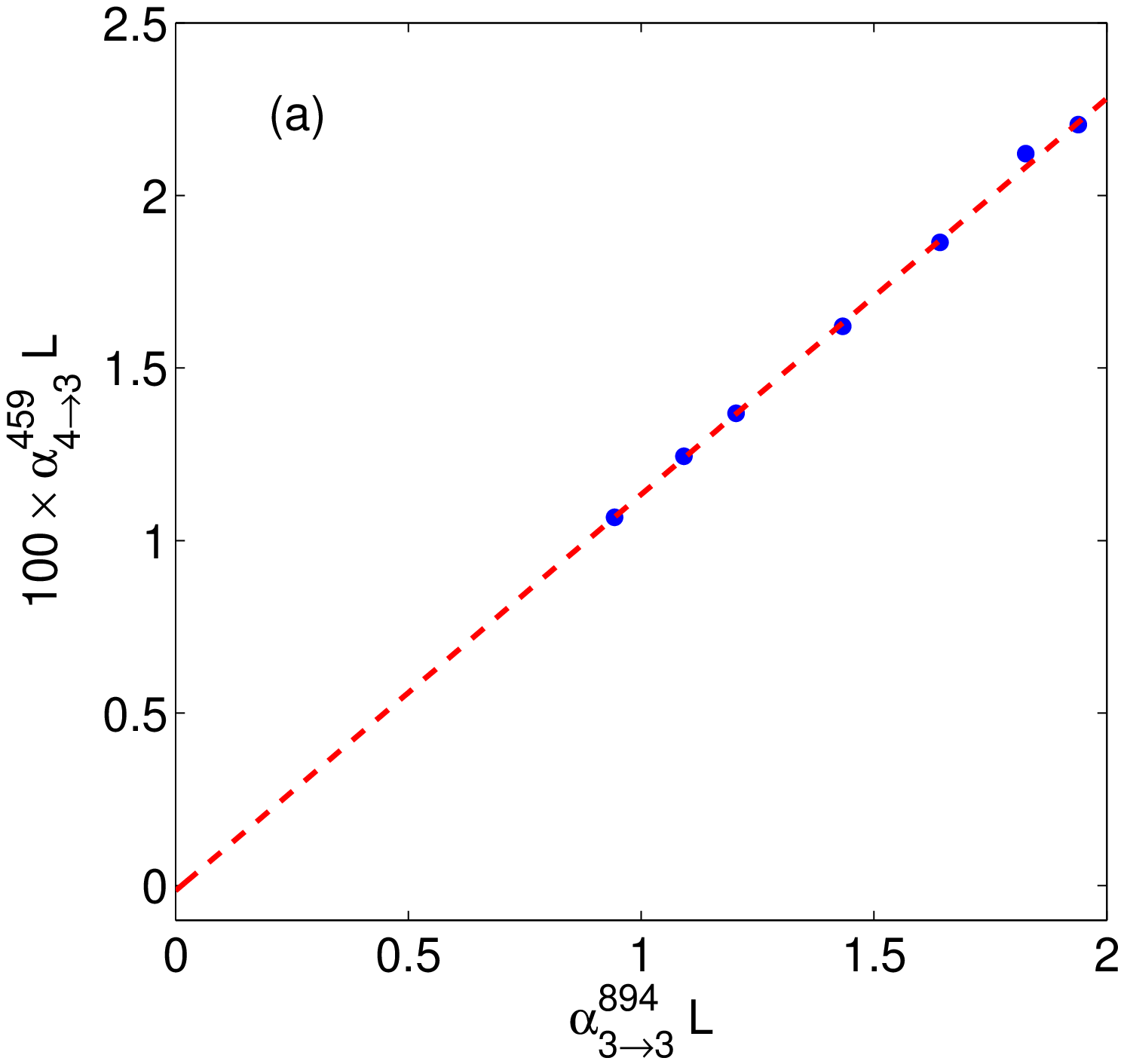}\\
  \includegraphics[width=7cm]{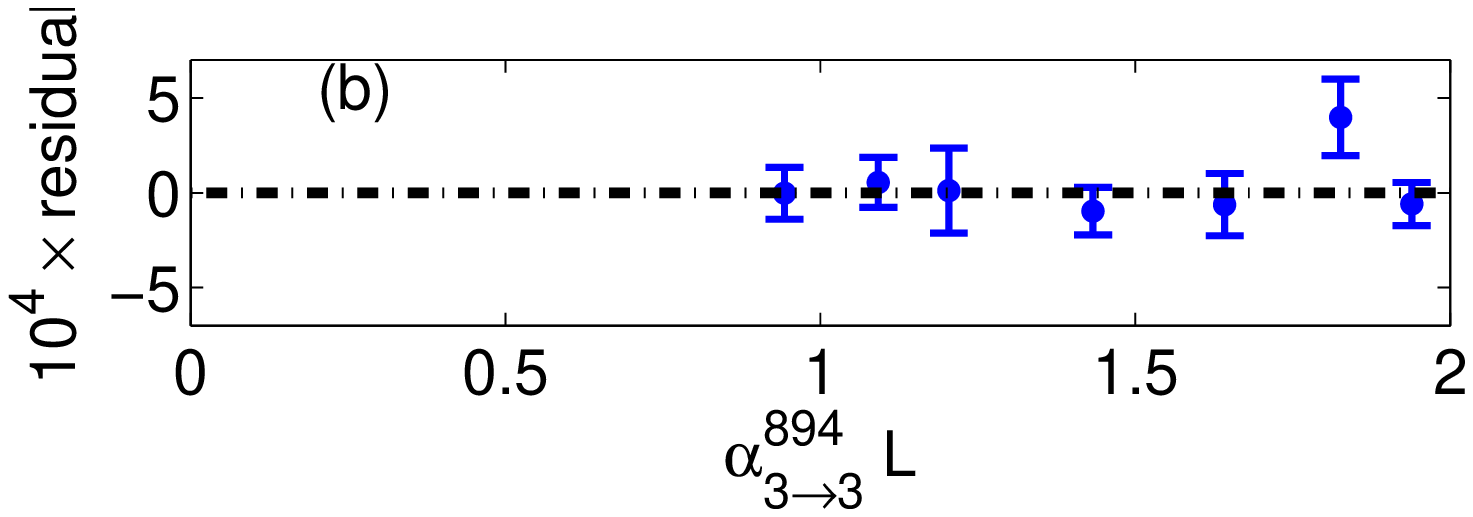}
  \caption{(Color on-line) (a) A plot the measured values of $\alpha_{4 \rightarrow 3}^{459} \: L$ at $\lambda$ = 459 nm versus $\alpha_{3 \rightarrow 3}^{894} \: L$ at $\lambda$ = 894 nm.  The red dashed line is the linear best fit to the data.  (b) The residual between the data points and the straight line fit.  The error bars on each residual indicate the standard deviation of the mean of the individual measurements.  }
  \label{fig:abs459vsabs894}
\end{figure}

In addition to the measurements described above using the $6s \: ^2S_{1/2}  \rightarrow 7p \: ^2P_{3/2} $ resonance at 456 nm, we have carried out similar absorption measurements on the $6s \: ^2S_{1/2}  \rightarrow 7p \: ^2P_{1/2} $ absorption resonances at $\lambda $ = 459 nm.  We show an example of a transmission spectrum on this line in Fig.~\ref{fig:absdata459}(a).
The absorption strength is rather weak on this line, resulting in larger relative fluctuations.  (We have also removed an amplifier with a gain of 6 from the detection electronics for these measurements, decreasing the overall magnitude of the signal, but not impacting the signal-to-noise ratio.)  We show the fit to the transmission spectrum with the red dashed line, and the residual difference between the measured transmission and the fitted curve in Fig.~\ref{fig:absdata459}(b). We measured four transmission spectra at each of seven cell temperatures, and plot $\alpha_{4 \rightarrow 3}^{459} \: L$ versus $\alpha_{3 \rightarrow 3}^{894} \: L$ in Fig.~\ref{fig:abs459vsabs894}(a).

\begin{table}
\begin{tabular}{|c|c|c|}
  \hline
     & $6s \: ^2S_{1/2} \: F=4  \rightarrow  $    & $6s \: ^2S_{1/2} \:  F=3  \rightarrow  $  \\
      &  \hspace{0.2in} $ 7p \: ^2P_{1/2} \: F=3$  &  \hspace{0.2in} $ 7p \: ^2P_{1/2} \: F=3$ \\ \hline \hline
  $m \times 10^{2}$ &  $1.148 \: (15)$ &  $0.3844 \: (91)$   \\
  $b \times 10^{5}$   &  $-15 \: (22) $ &  $-6 \: (11) $   \\
  reduced $\chi^2$   &  $1.02   $ & $1.52 $  \\
  $q_{J,F \rightarrow J^{\prime},F^{\prime}}$  &  7/4  &  7/12  \\
  $\langle 7P_{1/2} || r || 6S_{1/2} \rangle$   &  $0.2787 \: (18) $ & $0.2794 \: (33) $    \\ \hline
\end{tabular}
\caption{Results of the fit to the data for the 459 nm absorption line, including the slope $m$, the intercept $b$, the reduced $\chi^2$, and the reduced matrix element in atomic units $a_0$. }
\label{table:results459}
\end{table}
The weighted average of these determinations yields $\langle 7P_{1/2} || r || 6S_{1/2} \rangle = 0.2789 \: (16) \: a_0$, with a fractional uncertainty of 0.6\%, including the effect of the uncertainty of the $\langle 6P_{1/2} || r || 6S_{1/2} \rangle$ matrix element.  The decreased precision of this measurement, compared to that of the $\langle 7P_{1/2} || r || 6S_{1/2} \rangle$ moment, stems from the relatively smaller ratio of the absorption depth of this transition to that of the D$_1$ line.  The low absorption strength at 459 nm made reliable measurements at lower vapor cell densities difficult, while at higher densities,
we were less confident in the fits of the absorption profiles of the D$_1$ line (894 nm).  We suggest that a possible future solution to this problem would be to perform concurrent measurements of the absorption depth of the $6s \: ^2S_{1/2} \rightarrow 7p \: ^2P_{1/2}$ and $6s \: ^2S_{1/2} \rightarrow 7p \: ^2P_{3/2}$ transitions.  Unfortunately, this measurement would require two blue laser sources to assure that the vapor density in the cell is unchanged from one measurement to the other, and so we were unable to pursue this measurement further.

We compare our results to prior theoretical and experimental determinations in Table~\ref{table:ResultComparison}.
\begin{table}
\begin{tabular}{|r|c|c|}
  \hline
\multicolumn{1}{|c|}{group}    & $\langle 7P_{3/2} || r || 6S_{1/2} \rangle $ & $\langle 7P_{1/2} || r || 6S_{1/2} \rangle $    \\ \hline \hline
\multicolumn{1}{|l|}{\emph{expt.}} & & \\
   Shabanova {\it et al.}, Ref.~\cite{ShabanovaMK79}     & $0.583 \: (10)$  &  $0.2841  \: (21)$  \\
  Vasilyev {\it et al.}, Ref.~\cite{VasilyevSSB02}    & $0.5856 \: (50) $  & $0.2757 \: (20) $ \\
  This work                                         &  $  0.5780 \: (7) $ & $ 0.2789 \: (16)  $    \\
 & & \\
\multicolumn{1}{|l|}{\emph{theor.}} & & \\

   Dzuba {\it et al.}, Ref.~\cite{DzubaFKS89}   & 0.583  &  $0.275  $   \\
   Blundell {\it et al.}, Ref.~\cite{BlundellJS91}   & 0.576  &  $0.280  $   \\
   Safronova {\it et al.}, Ref.~\cite{SafronovaJD99}   & 0.576  &  $0.279  $   \\
   Derevianko {\it et al.}, Ref.~\cite{Derevianko00}   &   &  $0.281  $   \\
  Porsev {\it et al.}, Ref.~\cite{PorsevBD10}         &   &  $0.2769  $    \\ \hline
\end{tabular}
\caption{Comparison of the matrix elements determined in this work to those of prior experimental and theoretical works. All values are in units of $a_0$.}
\label{table:ResultComparison}
\end{table}
In comparing our measurement of $\langle 7P_{3/2} || r || 6S_{1/2} \rangle $ to that of Ref.~\cite{VasilyevSSB02}, we note that our uncertainty is smaller by a factor of $\sim$7, and that our results differ by 1.3\%, somewhat larger than the 0.8\% uncertainty of that result.  We note that our technique does not require a precise determination of the vapor density.  Our result is also in reasonable agreement of the most recently reported calculated value of Ref.~\cite{SafronovaJD99}.  For $\langle 7P_{1/2} || r || 6S_{1/2} \rangle $, our result is only of slightly higher precision than that of Ref.~\cite{VasilyevSSB02}.  The difference between these measurements of $\sim$1.1\% is somewhat larger than the stated uncertainties of the individual results.  A comparison with the most recent theoretical result~\cite{PorsevBD10} for this moment shows that our result is 0.7\% greater, while the result of Ref.~\cite{VasilyevSSB02} was 0.4\% smaller.

\section{6s - 7s Scalar and Vector Polarizability }\label{sec:alphabeta}
As discussed in the introduction, the vector polarizability $\beta$ for the 6s - 7s transition in atomic cesium has played a central role in parity non-conservation measurements.  Accurate direct calculation of this parameter, however, is difficult, as there is significant cancelation between various terms in the sum-over-states expression, such as Eq. (3) of Ref.~\cite{DzubaFS97}.  The two methods of determining $\beta$ indirectly provide somewhat different values.   The first, through calculation of $M_1^{\rm hfs} $~\cite{BouchiatG88,SavukovDBJ99,DzubaF00} and a precise measurement of the ratio $M_1^{\rm hfs} /\beta$~\cite{BennettW99}, yields $\beta =26.957 \: (51) \: a_0^3$~\cite{DzubaF00}.  The second uses a calculation of $\alpha$~\cite{DzubaFS97,VasilyevSSB02,DzubaFG02} and a precise measurement~\cite{ChoWBRW97} of the ratio $\alpha / \beta =  -9.905 \: (11) $ to find $\beta=27.15 \: (11) \: a_0^3$~\cite{DzubaFG02}.  These two values of $\beta$ differ by $ 0.19  \: a_0^3$, or 0.7\%, a difference that exceeds the stated uncertainties.  In this section, we use our new determination of the radial matrix elements to provide a correction to the scalar polarizability, and resolve, at least partially the discrepancy between these two values.

In the sum-over-states form of the scalar polarizability $\alpha$ presented in Eq.~(\ref{eq:alphaSumOverStates}), we have re-evaluated the contributions by the $7p \: ^2P_{3/2}$ and $7p \: ^2P_{1/2}$ intermediate states. Using $\langle 7P_{3/2} || r || 6S_{1/2} \rangle  = 0.5780 \: (7) \: a_0$ and $\langle 7P_{1/2} || r || 6S_{1/2} \rangle = 0.2789 \: (16) \: a_0$ in place of the similar quantities reported by Vasilyev {\it et al.}~\cite{VasilyevSSB02}, we have calculated a correction to $\alpha$ of $\Delta \alpha = 0.93 \: a_0^3$.  Since the uncertainty of the previous determination of $\alpha$ was dominated by the uncertainty of these two matrix elements, we have also re-evaluated the uncertainty, and find that it is reduced from 0.38\%~\cite{VasilyevSSB02} to 0.28\%.  We expect that this correction $\Delta \alpha$ can be applied to the results of Ref.~\cite{DzubaFG02} as well, since this determination also uses the experimentally determined radial matrix elements of Ref.~\cite{VasilyevSSB02}. Using the ratio $\alpha / \beta =  -9.905 \: (11) $~\cite{ChoWBRW97}, we determine a correction to $\beta$
of $\Delta \beta = -0.094 \: a_0^3$, which reduces $\beta$ from $27.15 \: (11) \: a_0^3$ to $27.06 \: (7) \: a_0^3$.  This value of $\beta$ is in better agreement with the value determined through $M_1^{\rm hfs} $, differing by $0.09 \: (9) \: a_0^3$, or 0.4\%, where the uncertainty in parentheses is the quadrature sum of the uncertainties determined by the two separate methods.  We also note that the uncertainty in the $\langle 7S_{1/2} || r || 6P_{3/2} \rangle$ matrix element is now the limiting factor in the sum over states evaluation of the scalar polarizability~\cite{VasilyevSSB02}.

\section{Conclusion}\label{sec:conclusion}
In this work, we have described our measurements of the absorption strength of the $6s \: ^2S_{1/2}  \rightarrow 7p \: ^2P_{3/2} $ the $6s \: ^2S_{1/2}  \rightarrow 7p \: ^2P_{1/2} $ lines in an atomic cesium vapor cell at $\lambda$ = 459 nm and $\lambda$ = 456 nm, respectively.  By simultaneously measuring the absorption strength of the D$_1$ line at $\lambda$ = 894 nm, and using the precisely determined transition matrix element for this line, we are able to determine $\langle 7P_{3/2} || r || 6S_{1/2} \rangle  = 0.5780 \: (7) \: a_0$ and $\langle 7P_{1/2} || r || 6S_{1/2} \rangle = 0.2789 \: (16) \: a_0$.  We have used these new determinations of these matrix elements to provide a new value of the vector polarizability for the $6s \: ^2S_{1/2}  \rightarrow 7s \: ^2S_{1/2} $ transition, $\beta = 27.06 \: (7) \: a_0^3$.  Our result for $\beta$ is in reasonable agreement with the value determined through the off-diagonal magnetic dipole moment.  We hope that this new determination of the radial matrix elements reported here stimulates an improved theoretical determination of the scalar polarizability $\alpha$, more complete than our estimates described in the previous section.

This material is based upon work supported by the National Science Foundation under Grant Number PHY-0970041.  We are also happy to acknowledge useful communications with M. S. Safronova.

\end{document}